%% file: paper.tex
\documentclass[letterpaper, 10 pt, conference]{ieeeconf}  

\IEEEoverridecommandlockouts                              

\usepackage{url}
\usepackage{booktabs} 
\usepackage{array}
\usepackage{subfig}
\usepackage{cite}
\usepackage{amsmath,amssymb,amsfonts}
\usepackage{graphicx}
\usepackage{textcomp}
\usepackage{xcolor}
\usepackage[ruled]{algorithm2e}
\usepackage{pifont}
\usepackage{tabularx,booktabs}
\usepackage{multirow}
\usepackage{makecell}

\newcommand{\isep}{\mathrel{{.}\,{.}}\nobreak}
\DeclareMathOperator*{\argmax}{arg\,max}

\newcommand{\change}[1]{\textcolor{black}{{#1}}}

\def\BibTeX{{\rm B\kern-.05em{\sc i\kern-.025em b}\kern-.08em
    T\kern-.1667em\lower.7ex\hbox{E}\kern-.125emX}}

\title{\LARGE \bf
A Novel Multi-Centroid Template Matching Algorithm and Its Application to Cough Detection*
}

\author{Shibo Zhang$^{1}$, Ebrahim Nemati$^{2}$, Tousif Ahmed$^{2}$, Md Mahbubur Rahman$^{2}$, Jilong Kuang$^{2}$, and Alex Gao$^{2}$
\thanks{*This work was supported by Samsung Research America.}
\thanks{$^{1}$Shibo Zhang is with Computer Science Department, Northwestern University. Work done during internship at Samsung Research America.
        {\tt\small shibo.zhang@northwestern.edu}}%
\thanks{$^{2}$Ebrahim Nemati, Tousif Ahmed, Md Mahbubur Rahman, Jilong Kuang, and Alex Gao are with Digital Health Lab, Samsung Research America, Mountain View, CA, USA.}%
\thanks{COPYRIGHT TRANSFER
The undersigned hereby assigns to The Institute of Electrical and Electronics Engineers, Incorporated (the "IEEE") all rights
under copyright that may exist in and to: (a) the Work, including any revised or expanded derivative works submitted to the IEEE
by the undersigned based on the Work; and (b) any associated written or multimedia components or other enhancements accompanying the Work.}
}

\begin{document}

\maketitle
\thispagestyle{empty}
\pagestyle{empty}

\begin{abstract}

Cough is a major symptom of respiratory-related diseases. There exists a tremendous amount of work in detecting coughs from audio but there has been no effort to identify coughs from solely inertial measurement unit (IMU). Coughing causes motion across the whole body and especially on the neck and head. Therefore, head motion data during coughing captured by a head-worn IMU sensor could be leveraged to detect coughs using a template matching algorithm. In time series template matching problems, K-Nearest Neighbors (KNN) combined with elastic distance measurement (esp. Dynamic Time Warping (DTW)) achieves outstanding performance. However, it is often regarded as prohibitively time-consuming. Nearest Centroid Classifier is thereafter proposed. But the accuracy is comprised of only one centroid obtained for each class. Centroid-based Classifier performs clustering and averaging for each cluster, but requires manually setting the number of clusters. We propose a novel self-tuning multi-centroid template-matching algorithm, which can automatically adjust the number of clusters to balance accuracy and inference time. Through experiments conducted on synthetic datasets and a real-world earbud-based cough dataset, we demonstrate the superiority of our proposed algorithm and present the result of cough detection with a single accelerometer sensor on the earbuds platform.
\newline

\indent \textit{Clinical relevance}— \change{Coughing is a ubiquitous symptom of pulmonary disease, especially for patients with COPD and asthma. This work explores the possibility and and presents the result of cough detection using an IMU sensor embedded in earables.}

\end{abstract}


\input{1intro.tex}

\input{2related.tex}
\input{4method.tex}

\input{5result.tex}

\input{6conclusion.tex}



\bibliographystyle{IEEEtran.bst}
\bibliography{paper.bib}

\end{document}

%% file: 1intro.tex
\section{Introduction}\label{sec:intro}

%
%
%

Time series problems are ubiquitous in many facets of human life, such as finance, biology, physiology, psychology, multimedia, etc. Time series classification has been intensely researched in machine learning community over decades.
%
%
A group of approaches extract a list of statistical features from time series data and then apply traditional classification algorithms on the features. These approaches can be viewed as processes of data compression, while at the same time also inherit the disadvantage of information loss. Because the feature engineering requires domain knowledge, and manually designed features often fail in capturing fine-grained details and sophisticated nuances in original time series, thus feature-based methods often give satisfactory results in relatively simple tasks and have limited performance on a larger range of applications.
In the most recent decade, deep learning based methods have gained rapid popularity in time series classification, due to its automatic feature extraction capability and superior performance when a large mount of training data is available. 
However, Convolutional Neural Network (CNN) is not capable of dealing with various length of input data, which impedes its usability in many scenarios. Recurrent Neural Network (RNN), although investigated intensively, such as equipped with various Long Short-Term Memory (LSTM) cells, is still relatively hard to converge during training which adds its difficulty to apply. 
Besides, the demanding requirement for large labeled datasets makes it unsuitable for many time series applications like wearable-based activity recognition. 
Another group of methods build generative models for targeted time series. One typical example is Hidden Markov Model (HMM). Unfortunately, as a generative model, HMM often gives less competitive classification performance.

Apart from aforementioned feature-based and model-based approaches, distance-based approaches are widely studied and applied in time series data. One simple but powerful classification approach is KNN combined with elastic distance metrics. 
DTW, as the most widely acknowledged time series measurement, is imported in clustering and classification tasks.
KNN-DTW method has an outstanding accuracy record in many literature \cite{DTW}. However, it is also featured by its time-consuming referencing process, which hinders its real-world applicability. 
The reference time is especially critical for real-time online applications.
The training process can take place offline or on a powerful device with data transferred, but the inference usually needs to be online on low-resource platform, such as wearable systems. 

Template matching is an efficient solution to the reference time issue. 
As a branch of distance-based approaches, it is well studied in domains of classification, data mining, information retrieval, etc~\cite{TDM}. 
A template matching method involves two components: template generation and template retrieval. Template generation focuses on how to generate the best template that is representative of the training data of interest; Template retrieval uses the generated template to search in the test data for the data that resembles the template the most. 
However, most of the existing works focus on the template retrieval part and only several aim at template generation~\cite{Chaoji2008}. 

\begin{figure}
\centering
\vspace{.3cm}\includegraphics[width=.4\textwidth]{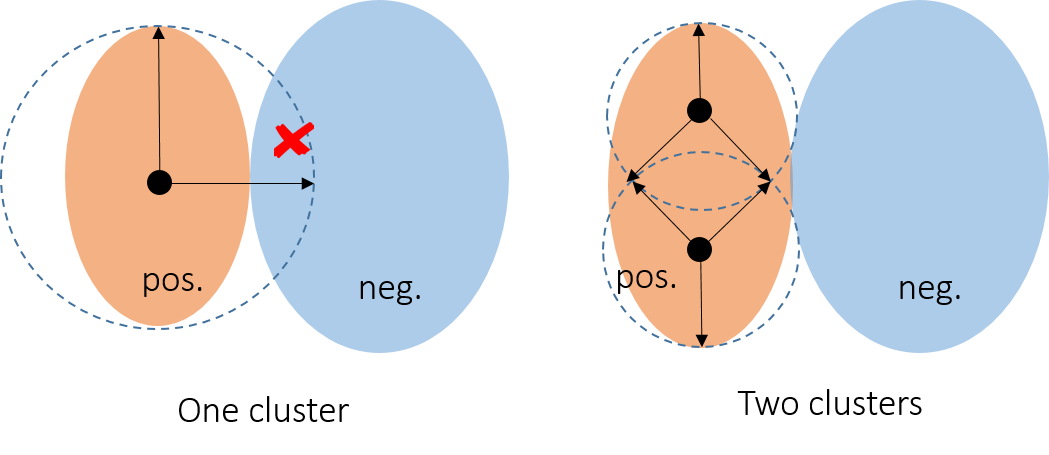}
\caption{Intuition of Multi-Centroid Classifier.}
\label{fig:intuition}
\end{figure}

In template matching, usually only the positive class, i.e., the shape of interest, is modeled, because the negative class can have a rather large range of possible shapes and wider distribution in measurement space. Therefore, negative samples often take little part in template generation. 
However, as illustrated in Fig.~\ref{fig:intuition}, 
we notice that the negative samples can actually guide the selection of the templates or clusters for positive class. 
We utilize the negative class samples when generating templates for positive class, which allows us to optimize the number of templates and achieve a superior classification performance.

In addition to inference speed, another requirement by embedded system is the model size. Since the availability of RAM is often limited when deployed on wearable devices, trained models often need additional model pruning to fit the embedded device.
In this work, we propose a novel fast and space-efﬁcient multi-centroid template-matching algorithm for time series classification with a minimal model size, which can potentially enable on-chip real-time data processing and pattern recognition.

\change{Coughs could be the sign of serious lung diseases, such as Chronic Obstructive Pulmonary Disease (COPD), asthma, pneumothorax, atelectasis, bronchitis, and even lung cancer. 
lung diseases yield staggering healthcare costs, bringing a huge economic and social burden to patients and the society. 
Since cough is an important indicator of respiratory-related disease, reliable detection of cough using wearables is especially desirable in healthcare community among doctors and clinical practitioners~\cite{ebrahim_embc2020}. } 

\change{In most recent years, automatic cough detection based on wearables has attracted a lot of attention from researchers. A large body of work has recently emerged, with wearable sensors showing promise in automatically identifying coughs and classifying different types of coughs. 
For example, researchers have proposed using smartphone microphone to detect coughs \cite{coughmic2011, listen2cough,Nemati_bodynets}. 
However, current technology for cough detection relies on audio recordings and requires rather long processing time~\cite{Nemati2020}. 
%
So far, there has been no effort to identify coughs from solely inertial measurement unit (IMU). Coughing includes a big inhale, a burst of air out of lung and a vocal phase as the tail. This causes motion across the whole body and especially on the neck and head. Therefore, the unique head motion signature caused when coughing could perhaps be captured to identify a cough event using a head-worn IMU sensor. 
In this work, we explore the possibility of detecting cough using an IMU sensor in earbuds for the first time.}

The remainder of this paper is organized as follows: In Section II, the related work on time series classification and other template matching techniques is presented and discussed. In Section III, the system architecture and algorithm design are presented. In Section IV, the experimental result is reported followed by conclusion in Section V.

%% file: 2related.tex
\section{Related Work}

\change{Because as far as we are aware, this is the first work aiming at head motion-based cough detection, in this section, we focus on discussing the existing literature on template matching and time series classification.}

\change{All types of} time series classification methods can be categorized into three classes: feature based, generative model based and distance based methods. 
Feature based methods either manually extract features from time series and leverage traditional instance-based classifiers, or automatically learn feature representation from raw data and conduct classification task (deep learning). Generative model based methods build generative models (eg. HMM and auto-regressive models) to model the time series generation process. Distance-based methods focus on measuring the similarity between instances and utilizing neighbors-based learning methods such as KNN classifier.

\change{KNN combined with elastic distance measurement (esp. 1NN-DTW) is shown to achieve unbeatable classification performance~\cite{DTW, Bagnall2016}. 
However, it is often regarded as prohibitively time-consuming. 
Nearest Centroid Classifier is thereafter proposed~\cite{NCC}. Unfortunately, with only one centroid obtained for each class, the accuracy is comprised as a result, although realistic running time is achieved.
Centroid-based Classifier is another invention, which performs clustering and averaging for each cluster~\cite{one_centroid}. It models both positive and negative classes, but requires manually setting the number of clusters. Both of the above classifiers are template matching approach, which utilizes template(s) produced from instances in the training set. 
Besides, many more types of time series classifiers are proposed to achieve improved accuracy~\cite{Bagnall2016, TSF, Grabocka2014, GORECKI201498, cote, ee}. 
For example, a collective of transformation-based ensembles (COTE), formed of 1-NNs with euclidean distance and/or dynamic time warping, is conclusively shown to have better accuracy than both of these approaches~\cite{cote}.
Elastic ensemble (EE) combines 11 constituent nearest neighbor classifiers that utilize variants of DTW and edit distance-based measures, including weighted and derivative DTW, longest common subsequence, edit distance with real penalty, and so on~\cite{ee}. Although ensemble classifiers significantly outperform baseline methods, the runtime is overlooked.
In this work, we emphasize on the runtime during inference while keeping a high accuracy.} 

Template matching approach is widely used in wearable studies. In~\cite{teethtap}, the authors adopted KNN method with DTW measurement to perform time series classification to recognize tooth tapping action for human-computer interaction. Holt et al. used KNN algorithm using various DTW variants in gesture recognition.~\cite{DTW_gesture}. Liu et al. adopted template matching algorithm in activity recognition and a gesture-based user authentication using a single three-axis accelerometer~\cite{4912759}. The authors in~\cite{image} also employ template matching approach to apply image-based gesture recognition. 
Discovery of highly representative templates is of significance to wearable studies, activity recognition studies and many more application domains in other fields.
%

%

In this work, we propose a novel self-tuning multi-centroid classifier, which can automatically adjust the number of clusters to balance accuracy and running time. 

%% file: 4method.tex
\section{Multi-Centroid Classifier (MCC)}\label{sec:method}

\subsection{Motivation}\label{sec:architecture}

%
The proposal of MCC is intuitive as depicted in Fig.~\ref{fig:intuition}. 
All the positive samples are clustered into one or more clusters. Each cluster outputs its centroid as a template along with a threshold which defines the boundary of the cluster. 
As shown in the left figure, when a single template is generated from positive samples, the selected threshold unfavorably includes part of negatives samples in the cluster. Thus, positive and negative samples cannot be accurately separated by a single threshold. 
If we use two clusters to produce two templates, positive and negative samples can be separated accurately by a threshold for each cluster. Note that as more clusters provide better accuracy, running time is increasing as a trade-off. 
In order to automatically learn the best number of clusters (templates) to optimize accuracy and running time, we propose this procedure: it starts from only one cluster and then iteratively splits the cluster with the worst performance into two clusters to continuously obtain a higher accuracy, until a desired accuracy is obtained. In this way, the algorithm automatically learns the best number of clusters (templates) which leads to achieving a high accuracy. 
Our proposed MCC contains three main components: cost function, discrepancy-based clustering, and cluster averaging. \change{We will explain each of them in details in this section.}

\begin{figure}
\centering
\vspace{.3cm}\includegraphics[width=.45\textwidth]{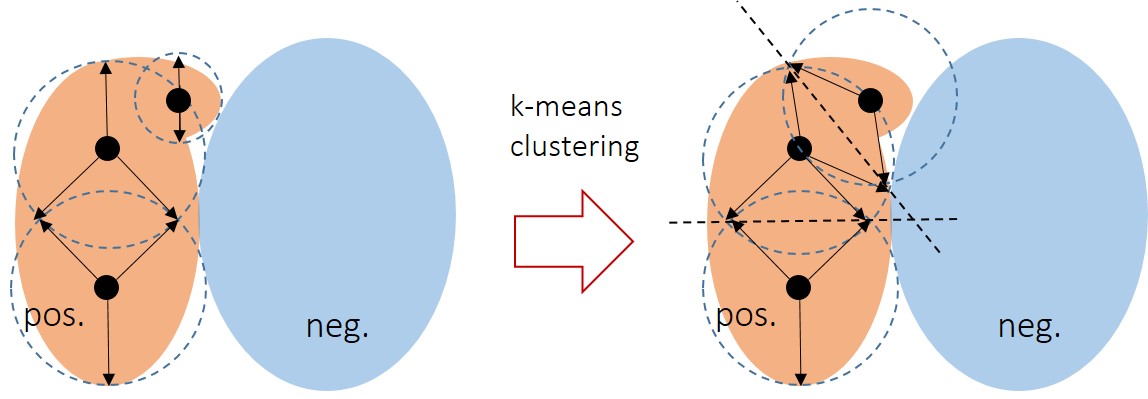}
\caption{Illustration of the effect if using k-means in MCC. The left shows an ideal condition for template generation. The right shows the clustering iteration will not stop at the ideal condition leading to non-optimal result.}
\label{fig:kmeans}
\end{figure}

\subsection{Discrepancy Cost}\label{sec:cost}
To evaluate the performance of clusters in each iteration, we propose a novel discrepancy cost for MCC to measure the performance of clusters. 
We denote the training dataset as $\{\mathbf{X}, \mathbf{Y}\}$. $\mathbf{X}$ represents the attributes and $\mathbf{Y}$ the labels. We use $\mathbf{X_P}$ to represent the attributes of positive samples and $\mathbf{X_N}$ the attributes of negative samples. 
Assume there are $M$ positive samples and $N$ negative samples, and the training stage has $T$ iterations. 
In $t$-th iteration, there are $K_t$ clusters in total. 
The centroid of cluster $k$ is denoted by $C_k$. 
Cluster $k$, $k=\{1\isep K_t\}$, with boundary defined by its threshold, has $M_k$ positive samples and $N_k$ negative samples.
We use $m_{ik},  i=\{1\isep M_k$\} to represent the positive sample in cluster $k$, and $n_{jk}, j=\{1\isep N_k$\} for the negative sample in cluster $k$. 
In each iteration, we require every positive sample to be assigned to one (or more) cluster, that is
\begin{equation}
\begin{split}
\forall~ t=\{1\isep T\}, ~ \bigcup_{k=1}^{K_t} \{m_{ik} | i=1\isep M_k\} = \mathbf{X_P}
\end{split}
\end{equation}

For simplicity, we omit $t$ in the following analysis. 
We hope $N_k=0$ under an ideal condition. Before arriving at a perfectly accurate classifier, we want to minimize the number of negative samples and maximize the number of positive samples in cluster $k$. Besides, we want the positive samples in cluster $k$ to be closer to the centroid $C_k$ than the negative samples in cluster $k$. 
We calculate the distance between each sample in cluster $k$ and the centroid $C_k$, then we get $M_k + N_k$ distances, we hope the $M_k$ positive samples’ distances $d(m_{ik}, C_k), i=\{1\isep M_k$\} are all smaller than the $N_k$ negative samples' distances $d(n_{jk}, C_k), j=\{1\isep N_k$\}. That indicates if we calculate $d(m_{ik}, C_k) - d(n_{jk}, C_k)$, which is the difference between each positive sample's distance and each negative sample's distance (in total $M_k\times N_k$ items),
we want the $M_k\times N_k$ differences to be all negative values. Thus, if we consider all the clusters and calculate the sum of all the positive differences (while ignoring all the negative values) as the cost function, the goal is to minimize cost function $L$
%
%
%
\begin{equation}
    {L = \sum_{k=1}^{K} \sum_{i=1}^{M_k} \sum_{j=1}^{N_k} h (d(m_{ik}, C_{k}) - d(n_{jk}, C_{k}))}
    \label{eq:KDE}
\end{equation}

\begin{equation}
    h(x) = \begin{cases}
    x & \text{ if } x>0 \\ 
    0 & \text{ if } x\leqslant 0 
    \end{cases}
\end{equation}


\subsection{Discrepancy-based Clustering}

\begin{figure}
\centering
\vspace{.3cm}\includegraphics[width=.3\textwidth]{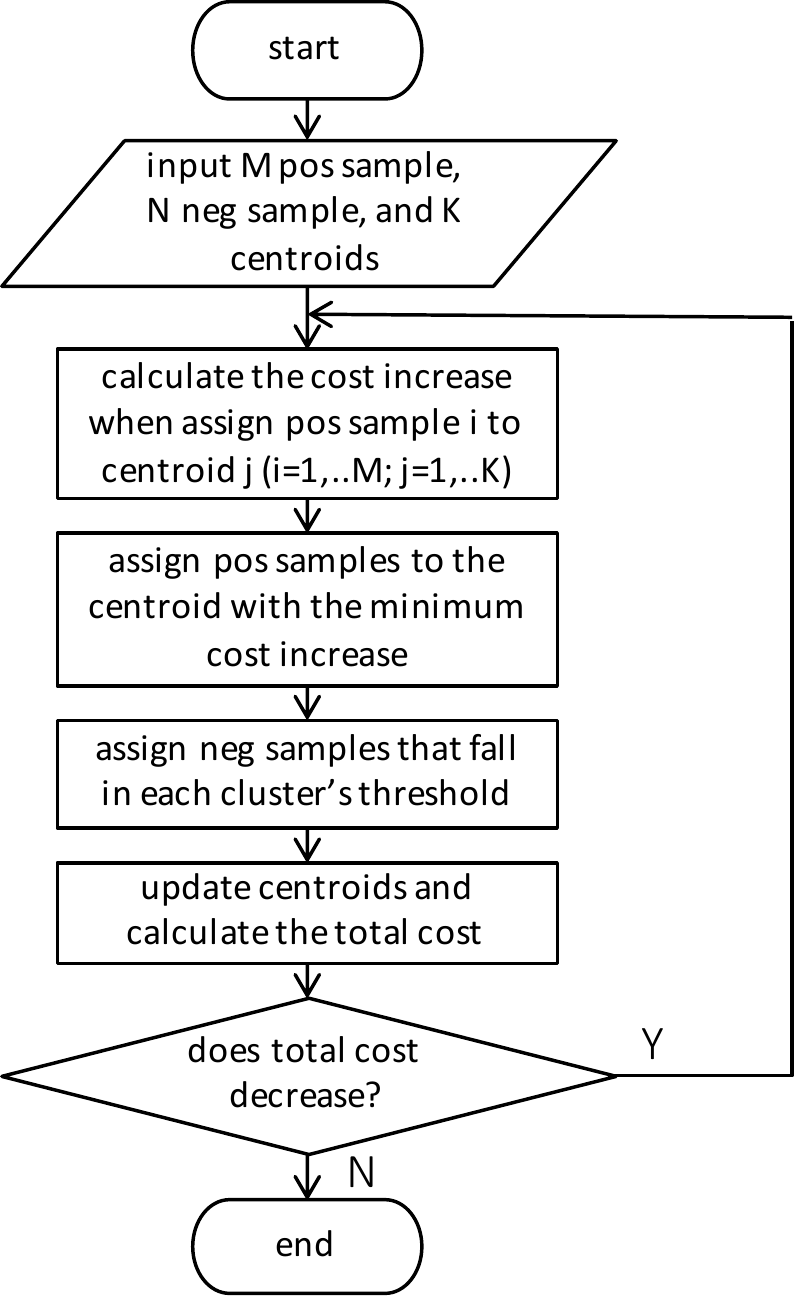}
\caption{\change{Flowchart of discrepancy-based clustering.}} 
\label{fig:discrepancy-based clustering}
\end{figure}

We assume the samples are perfectly assigned to clusters as shown in the Fig.~\ref{fig:kmeans} left. However, if we use k-means method, it will assign samples like in Fig.~\ref{fig:kmeans} right.
This goes against our ‘coarse-to-fine’ style divide-and-conquer principle. It will lead to more clusters than needed and non-optimal results.
From another angle, since each cluster centroid has its own threshold, thus deciding which cluster a specific sample belongs to cannot depends on a universal metric (Euclidean distance in above figure).
We formalize the requirements:
First, the assignment of a sample must not be determined by a single measurement (as in k-means). 
Second, the objective is to minimize the overall cost of all the clusters.
To cluster samples, we propose discrepancy based clustering.
Figure~\ref{fig:discrepancy-based clustering} shows the flowchart of discrepancy-based clustering which is used during the training of MCC. 
%

%
%



\subsection{Cluster Averaging} 
The averaging method should be capable to measure the similarity of the data. For non-time series data, Euclidean distance is a simple choice. 
For time series dataset, we adopt DTW Barycenter Averaging (DBA) method~\cite{Petitjean2011-DBA} in our algorithm because it reportedly provides outstanding averaging performance and is widely adopted in time-series clustering and averaging tasks. The choice of cluster averaging is not limited to Euclidean or DBA, but can be customized according to the specific application of MCC.



\subsection{Overall Algorithm}

\change{We present the training steps of MCC in Algorithm \ref{alg}. The input is the training dataset and a stop criterion that we need to set. We initiate the iterations from one cluster and a randomly selected seed centroid from the positive samples. Then we perform discrepancy-based clustering to obtain the total cost and updated centroid, along with the threshold. 
If the total cost is larger than the stop criterion, we randomly select two seed centroids from the positive samples in the only one cluster and repeat the discrepancy-based clustering with new centroids. After the clustering step, we update the centroids, cost for each cluster, and the total cost. Then if the total cost still exceeds the criterion, we repeat the procedure until it is less than or equal to the criterion, with the number of clusters increased by one in each iteration. 
How to choose the stop criterion seems trivial. However, the stop criterion can be set using training accuracy, then at the end of each iteration, we calculate the current training accuracy and compare with the criterion, with other steps kept unchanged. 
}

\begin{algorithm}[]
\vspace{.1cm}\SetAlgoLined
    \SetKwInOut{Input}{input}
    \SetKwInOut{Output}{output}
    \Input{Positive and negative training samples and stop criterion $H$\\}
    \Output{$K$ templates $C_k, k=1, 2, \dots K$ and $K$ thresholds}
 Initialize number of clusters $K=1$; \\
 Assign all the positive samples to cluster $R_1$; \\
 Randomly select seed centroid $C_1$ from positive samples of cluster $R_1$; \\
 Do discrepancy-based clustering with $C_1$ to obtain total cost $L$, updated centroid $C_1$ and threshold;\\
 \While{total cost $L$ $>$ stop criterion $H$}{
  Select $R_t$ with the highest cost $t = \argmax_i L_i$;\\
  Randomly select two seed centroids from the positive samples of $R_t$;\\
  Do discrepancy-based clustering using $K+1$ centroids $C_1, \dots, C_{K+1}$ to obtain total cost $L$, updated $K+1$ centroids, and $K+1$ thresholds;\\
  Calculate cost $L_k$ for each of the $K+1$ clusters;\\
  $L$ = $\sum_{k=1}^{K+1} L_k$;\\
  $K = K + 1$;\\
 }
 \caption{Training Multi-Centroid Classifier}\label{alg}
\end{algorithm}

\change{
The inference steps are straightforward. 
During inference, we calculate the distance between the test sample and each centroid, and the distance is compared with the corresponding threshold. If the distance is smaller than the threshold, then the prediction of the test sample is positive. If the distance is larger, then the test sample is negative.}

In this section, we propose a generic novel method which can balance the trade-off of running time and accuracy. Using a divide-and-conquer paradigm, it recursively breaks down a problem into two or more sub-problems until these become simple enough to be solved directly. Our method can automatically select the number of clusters. When the number of clusters is the number of positive samples, it deduces to an approach similar to 1NN-DTW method (1NN compares a test instance with both positive and negative instances while ours only compares with positive samples). Compared with nearest centroid classifier using classical clustering methods, it can take advantage of negative samples to achieve a better classification result. 




%% file: 5result.tex
\section{Experimental Result}\label{sec:res}
We conducted experiments on three datasets to validate the efficacy and efficiency of our proposed method and \change{explore the possibility of IMU-based cough detection}.


\subsection{Synthetic Dataset}

\begin{figure}
\centering
\hspace{-.5cm}\includegraphics[width=.3\textwidth]{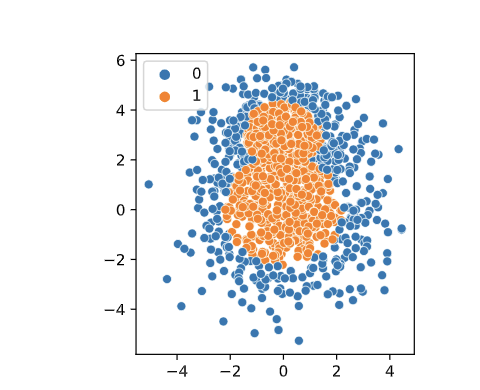}
\caption{Synthetic snowman dataset.}
\label{fig:snowman}
\end{figure}

We created a synthetic “snowman” dataset by combining two Gaussian distributions ($\mu_1$=$(0,0)$, $\sigma_1$=$3$; $\mu_2$=$(0,3)$, $\sigma_2$=$1$) as shown in Fig.~\ref{fig:snowman} to validate our algorithm. 
The dataset has the size of 1334 samples, each of which has two non-time series features. The positive to negative ratio is about 2:1. 
We randomly select 70\% of the samples as training set and the left as test set. 
We employed this dataset to make a sanity test and compare the inference speed and model size of MCC with a large range of popular classifiers\footnote{All implemented in Python 3.7. Hardware platform: Intel i7-7500U 2.70GHz, 2 cores, 16GB RAM.}. 
Euclidean distance is used here in MCC algorithm.
Note that the comparison of inference time is actually not fair for MCC, since MCC implementation code has not been optimized for speed, in contrast to other highly-optimized algorithm implementations from a popular library (scikit-learn~\cite{sklearn} uses Cython and C functions for speedup). 

We compared MCC with ten popular traditional classifiers including KNN (K=1), KNN (K=3), Linear SVM, Radial Basis Function (RBF) SVM, Decision Tree (DT), Random Forest (RF), Quadratic Discriminant Analysis (QDA), AdaBoost, Naive Bayes, and Neural Network (NN) with 1 hidden layer and 100 neurons. All the baselines are imported from scikit-learn library~\cite{sklearn}. We saved each trained model locally after training. For MCC, we saved both the classifier itself and the obtained centroids with thresholds. 

When implementing MCC, the cost threshold can be replaced by accuracy threshold, which allows direct manipulation of desired test accuracy (if the training set and test set have identical distribution). 
It is such an advanced and practically useful feature that we can adjust the accuracy threshold to balance the accuracy and inference time.
We tested MCC classifier with two different accuracy thresholds: MCC (T1) with 0.9 threshold and MCC (T2) with 0.95 threshold. 
MCC (T1) classifier gained 90.1\% training accuracy with two centroids converged, and test accuracy of 95.9\%. MCC (T2) has 96.9\% training accuracy with 14 centroids, and the test accuracy is 99.2\%.

\begin{figure}
\centering
\vspace{.1cm}\includegraphics[width=.497\textwidth]{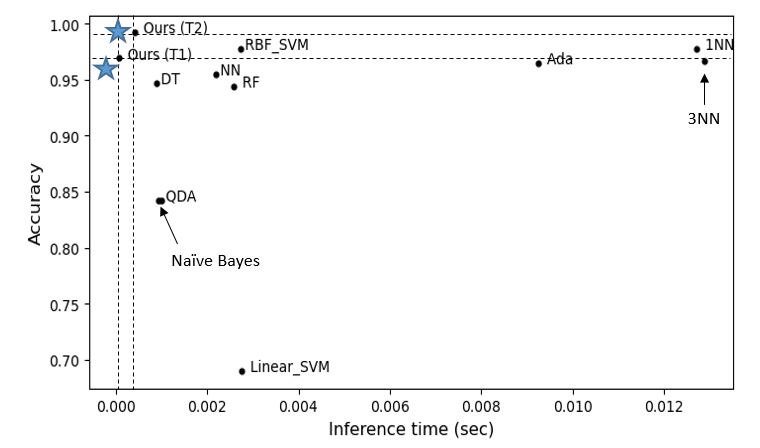}
\caption{Comparison of the accuracy V.S. inference time of different models on synthetic dataset.}
\label{fig:speed}
\end{figure}

\begin{figure}
\centering
\includegraphics[width=.48\textwidth]{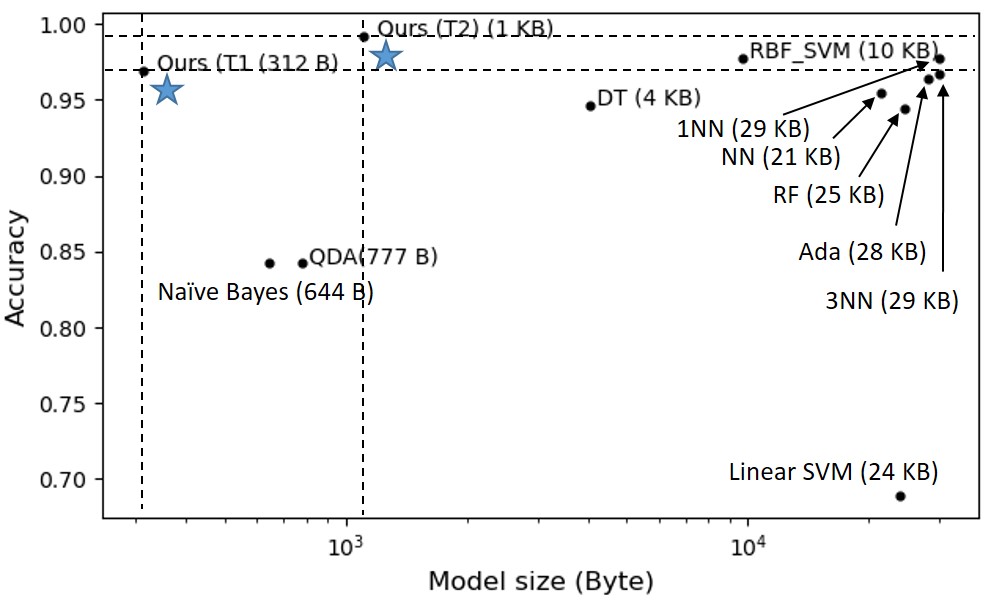}
\caption{Comparison of accuracy V.S. model size of different models on synthetic dataset.}
\label{fig:model size}
\end{figure}

After repeating each method for 1000 times, we show the averaged inference time V.S. accuracy in Fig.~\ref{fig:speed}. 
\change{MCC (T1) has an inference time of 0.062 ms and MCC (T2) takes 0.42 ms on average. Both of our MCC classifiers have shorter inference time than any of the baseline methods. In terms of test accuracy, MCC (T2) has achieved the best score (99.2\%), even better than KNN (K=1) due to its superior robustness, i.e., resistance to noise.} 
We then present the accuracy V.S. model size in Fig.~\ref{fig:model size}. It can be observed that our classifier provides a superior performance in aspect of model size, which is of important significance to low-resource embedded systems. It's worth noticing that our model can also be further optimized to be more space-efficient. 



\subsection{UCR Trace Dataset}


We used UCR Trace dataset~\cite{UCRArchive} to test its usability and inference speed in times series classification problems. The dataset size is 200, and the time series length is 275. There are 4 classes in the original dataset. We combined class 2 and 3 as training set and class 1 and 4 as test set. The positive and negative ratio is 1:1. We made a 7/3 ratio train/test set random split.
Since it is acknowledged that 1NN-DTW provides very accurate performance for time series classification problems, we use 1NN-DTW (implementation from tslearn library~\cite{tslearn}) as our baseline. The test accuracy is 100\%, and inference time is 2.1 s. The saved model size is 45 KB.
Our MCC finished at 2 clusters discovered and also a 100\% accuracy is achieved. The inference time is 0.011 s, which is 190 times shorter than baseline.
The model size is 1 KB, which is 45 times smaller than baseline, when we save both the templates (with threshold) and the model.

\subsection{Application to Earbud IMU-based Cough Detection}
\label{sec:imu_classifier}
Earbuds, as a sensing platform have been studied for human activity detection~\cite{teethtap}. 
We employed an earbud-based cough detection dataset as a test benchmark to evaluate the applicability of our classifier in wearable applications. 
Utilizing head motion data captured by accelerometer embedded in an earbud for cough detection can be formed as a template matching problem. 
Thus, our algorithm, which tries to find the optimal set of templates, is suitable for this problem.

\subsubsection{Data Collection}
We collected an earbud-based cough detection dataset which contains IMU data and audio during cough and non-cough sessions\footnote{\change{The experimental procedures involving human subjects described in this paper were approved by the Institutional Review Board.}}.
During these sessions, one earbud was worn by participants to collect accelerometer data and audio. 
In this work, we included 10 participants enrolled in the study and utilized the 3-axis accelerometer data to explore the possibility of cough detection using only accelerometer data. 
%
During data collection, in each cough session, 
We asked each participant to cough continuously for a period of time with a short break between every two coughs. 
Cough sessions include cough during seated (30 sec); cough during listening to music (30 sec), cough during yoga in a quiet environment (30 sec), cough during yoga in a noisy environment (30 sec), and cough during walking (30 sec). Non-cough activities include scripted speech (1 min), laughing (30 sec), free head motion while talking (30 sec), drinking (30 sec), and eating (30 sec). 

In total there are 651 coughs collected from 10 participants, of which 108 coughs take place in the seated session. On average there are 13 coughs captured in each cough session. We aim at detecting each single cough in every session while preventing false alarms.
%


\subsubsection{Data Processing}

\change{We preprocessed the 3-axis accelerometer data collected at 50 Hz using an average smoothing filter with 10 sample window size and a Butterworth high-pass filter with 1.5 Hz cut-off frequency.}
%
%
To gain the ground truth, we annotated the start and end of each cough with the aid of audio recordings \change{at first}. Because we observed a 0 to 300 ms time drifting between accelerometer and audio within each session, in order to obtain exact cough moments, we then labeled the moment of the most salient movement in accelerometer data as the center of a cough.
%
After data annotation, we sliced cough data using a 0.4 sec window with the annotated cough moment as the center. 
For non-cough sessions, we segmented the accelerometer data using a 0.4 sec sliding window with 0.1 sec stride size.


\subsubsection{\change{Model Development and Evaluation}}
We performed Leave-One-Subject-Out to evaluate our approach. 
%
During training, we used annotated cough data in \change{solely seated} cough sessions as positive instances. We randomly subsampled 500 segments from non-cough sessions as negative instances, in order to balance the ratio of positive/negative instances. \change{Because each of the 3 axes performs a different role in depicting the head motion during coughs, we use the processed data of 3 axes instead of combining them by calculating the magnitude. For simplicity}, we built a separate MCC model for each axis using DTW distance measurement.
When testing, we applied the trained models on the accelerometer data of the test participant's cough and non-cough sessions, with 0.4 sec window size and a stride size of one sample. Because we trained a single model for each axis, during testing we did a majority vote to combine the prediction of each axis. Then we aggregated and merged all the predicted cough windows to obtain final cough event prediction. No filtering was applied on the predicted label series.
%
%

We adopted sensitivity and specificity to evaluate the result. 
For cough sessions, we compared the predicted cough segments against the ground truth cough segments we obtained from audio recordings, ignoring the minor time-synch issue. True positives are defined as predicted cough segments intersected by cough ground truth. We calculated the sensitivity of cough sessions as a ratio of true positives to ground truth coughs. 
We used window-level specificity of non-cough sessions to test how well it can specify non-cough events. Due to the inconvenience of defining non-cough events, we define the specificity of a session as the ratio of falsely predicted cough windows to all the windows this session contains.

\subsubsection{Result}

\begin{figure}
\centering
\includegraphics[width=.25\textwidth]{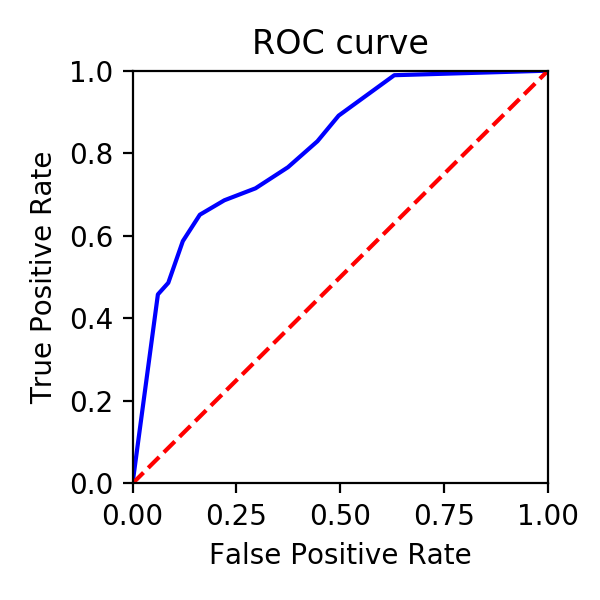}
\caption{Cough Detection ROC curve.}
\label{fig:roc}
\end{figure}

After training an MCC for each fold, on average 38.8 templates were generated. 
We received an average sensitivity of 65.1\% on seated cough session, and an average sensitivity of 56\% on all cough sessions. The average specificity is 83.8\% across all non-cough sessions for all the participants. 
A benefit of MCC is that we can manually adjust the trade-off between sensitivity and specificity. 
When we increase (decrease) the thresholds of all the templates, our model leans towards a higher sensitivity (specificity). 
We show the results of adjusting the thresholds in Table \ref{tab1} and present the ROC curve in Fig.~\ref{fig:roc}. 

\begin{table}[htbp]
\caption{Leave-One-Subject-Out Cough Detection Result}
\begin{center}
\begin{tabular}{|c|c|c|c|}
\hline
\textbf{\thead{Threshold \\ Adjustment}} & \textbf{\thead{Sensitivity \\ (Seated)}}& \textbf{\thead{Sensitivity \\ (All)}}& \textbf{{Specificity}} \\
\hline
-30 & 0.30 & 0.37 & 0.97 \\
\hline
-15 & 0.46 & 0.45 & 0.94 \\
\hline
0 & 0.65 & 0.56 & 0.84 \\
\hline
15 & 0.77 & 0.67 & 0.63 \\
\hline
30 & 0.89 & 0.76 & 0.50 \\
\hline
\end{tabular}
\label{tab1}
\end{center}
\end{table}



%

In the first two experiments of this section, we have proven the advanced runtime and accuracy of our algorithm. Here, we showcase the applicability of our method in a wearable study that utilizes only an accelerometer in an earbud to detect single coughs. 
As far as we know, this is the first work that shows recognition performance of cough and other confounding daily activities using only one accelerometer. 
We firmly believe that a better detection performance can be achieved with sensor fusion combining motion data and audio recording, and our accelerometer-based method can be a valuable component as one stage in a multi-modal sensor fusion pipeline. Besides, based on the method we developed, a personalized model could provide better performance than generalized model.

%% file: 6conclusion.tex
\section{Conclusion}\label{sec:conc}


We present a novel self-tuning multi-centroid template-matching algorithm, which provides outstanding accuracy and runtime, as well as the capability to automatically balance the accuracy and inference time. By performing experiments conducted on three datasets including a real-world earbud IMU-based cough detection dataset, we demonstrate the usability, efficacy and efficiency of our proposed algorithm. 
Besides, to the best of our knowledge, this is the first work that presents the recognition performance of cough and confounding daily activities using only one accelerometer in earbuds. 
We show huge potential of our proposed novel classification algorithm and the applicability to wearable-based human activity recognition and health monitoring scenarios.